\documentclass[aps,twocolumn,floatfix,amsmath,amssymb,showpacs,nofootinbib]{revtex4-1}

\usepackage{times,amsmath}
\usepackage{graphicx}
\usepackage{dcolumn}
\usepackage{bm}
\usepackage{float}
\usepackage{mathrsfs}
\usepackage{xcolor}
\usepackage[colorlinks,linkcolor=blue!80!black,urlcolor=purple!50!black,citecolor=green!50!black]{hyperref}

\begin{document}

\title{Dynamic Nuclear Polarization from Topological Insulator Helical Edge States}
\author{Antonio Russo}
\author{Edwin Barnes}
\author{Sophia E. Economou}
\affiliation{Department of Physics, Virginia Polytechnic Institute and State University, Blacksburg, Virginia 24061, USA}
\date{\today}

\def\eqnref#1{Eq.~(\ref{#1})}
\def\eqnsref#1#2{Eqs.~(\ref{#1}) and (\ref{#2})}
\def\figref#1{Fig.~\ref{#1}}
\def\secref#1{Sec.~\ref{#1}}
\def\tableref#1{TABLE~\ref{#1}}
\def\ket#1{\left|#1\right\rangle}
\def\bra#1{\left\langle#1\right|}
\def\braket#1#2{\left\langle#1\middle|#2\right\rangle}
\def\lA{\left|} \def\rA{\right|}
\def\lB{\left[} \def\rB{\right]}
\def\lC{\left\{} \def\rC{\right\}}
\def\lN{\left .} \def\rN{\right .}
\def\lP{\left(} \def\rP{\right)}
\def\fun{\lC\begin{array}{llcr}}
\def\fund{\end{array}\rN}

\begin{abstract}
Topological insulators are promising for spintronics and related technologies due to their spin-momentum-locked edge states, which are protected by time-reversal symmetry.  In addition to the unique fundamental physics that arises in these systems, the potential technological applications of these protected states has also been driving TI research over the past decade.  However, most known topological insulator materials naturally contain spinful nuclei, and their hyperfine coupling to helical edge states intrinsically breaks time-reversal symmetry, removing the topological protection and enabling the buildup of dynamic nuclear spin polarization through hyperfine-assisted backscattering.  Here, we calculate scattering probabilities and nuclear polarization for edge channels containing up to $34$ nuclear spins using a numerically exact analysis that exploits the symmetries of the problem to drastically reduce the computational complexity.  We then show the emergence of universal scaling properties that allow us to extrapolate our findings to vastly larger and experimentally relevant system sizes.  We find that significant nuclear polarization can result from relatively weak helical edge currents, suggesting that it may be an important factor affecting spin transport in topological insulator devices.
\end{abstract}
\maketitle

\section{Introduction}
Topological insulators (TIs) have a bulk band gap and gapless surface states due to the topologically nontrivial character of the occupied bulk bands \cite{Kane_2005,Kane_2005_2,Hasan_2010,Qi_2011}.  In the context of, e.g., HgTe \cite{Konig_2007,Bernevig_2006}, Bi$_2$Se$_3$ \cite{Fu_2007,Hsieh_2009,Zhang_2009,Hasan_2010}, the strong spin-orbit coupling leads to spin-momentum locking that makes the helical edge or surface states robust to disorder provided time-reversal symmetry is preserved \cite{Kane_2005,Xu_2006}.  However, most topological insulators have significant fractions of isotopes with spinful nuclei, and their hyperfine coupling to the edge or surface states violates time-reversal symmetry, in principle destroying the topological protection \cite{DelMaestro_2013,Tarasenko_2016,Hsu_2017,Bozkurt_2018}.  For example, Hg isotopes have 30\% spinful nuclei, while those of Te are 8\% spinful \cite{Schliemann_2003}.  While this effect could be mitigated by isotopic purification in some cases, this is not an option in others.  For example, there is only one isotope of Bi, and it has nuclear spin $9/2$.  Indeed, a recent experiment on Bi${}_2$Te${}_2$Se demonstrated a long-lived (on the order of days) spin memory attributed to dynamic nuclear polarization (DNP) \cite{Tian_2017}.  A full understanding of the microscopic mechanism behind this striking effect is currently lacking.

It is difficult to microscopically describe the process of DNP because of its many-body nature---the dimension of the Hilbert space scales exponentially with the number of nuclear spins.  This has made the study of electron-nuclear spin dynamics challenging in a variety of contexts, including semiconductor quantum dots \cite{Fischer_2009,Koppens_2008,Schliemann_2003,Khaetskii_2002,Petta_2005,Barnes_2011,Economou_2014,Yao_2006,Cywinski_2009,Coish_2010,Barnes_2012}, quantum wires \cite{Stano_2014,Mourik_2012,Pribiag_2013}, and more recently in transition metal dichalcogenides \cite{Sharma_2017}, generally necessitating the use of approximate methods \cite{Tanaka_2011,Lunde_2013} or very small systems.

In this paper, we consider two-dimensional topological insulators with spinful isotopes such as HgTe and obtain microscopic scattering results for one-dimensional spin-momentum-locked states interacting with nuclear spins (or other magnetic impurities) by exploiting the symmetries associated with helical edge states and the hyperfine interaction, dramatically reducing the computational resources needed.  We obtain exact scattering state solutions for up to $N=34$ nuclear spins.  In addition, we uncover universal scaling behavior that allows us to extrapolate our findings to much larger numbers of nuclear spins, enabling predictions for the buildup of dynamic nuclear polarization as a function of the edge current for realistic systems.

This paper is organized as follows. In \secref{sec:model} we introduce the model and discuss its symmetries. In \secref{sec:single} we focus on solving the single-nuclear-spin case and show how the symmetries constrain scattering parameters. In \secref{sec:multi} we present a scattering formula for arbitrarily many nuclear spins and solve it numerically for up to 34 spins. In \secref{sec:scaling} we find a scaling relation that allows us to extrapolate our results to arbitrarily many nuclei, and we use this to compute the DNP for a mesoscopic system. In \secref{sec:conclusion} we present our conclusions.  Detailed calculations and symmetry analyses are presented in the Appendices.

\section{Model\label{sec:model}}
Spin-momentum locked systems support both bulk and edge modes.  Remarkably, the edge modes can exhibit a Dirac-like dispersion \cite{Bernevig_2006,Hasan_2010}, in e.g., HgTe, and such systems are therefore governed by the Dirac Hamiltonian,
\begin{equation}
H(x) = -i\hbar v_0 \partial_x \sigma_z + \sum_{n=0}^{N-1} H_n^\text{HF}(x), \label{full-H}
\end{equation}
where $v_0$ is the effective electron velocity, and $\sigma_i$ are electron Pauli spin matrices.  This low-energy description assumes that the edge modes involved do not significantly hybridize with the bulk modes.  The first term captures the spin-momentum locking: a spin-up electron carries positive $x$ momentum.  The microscopic interactions $H_n^\text{HF}$ couple the electron and nuclear spins.  Ref.~\cite{Bernevig_2006} used a $\mathbf{k}\cdot\mathbf{p}$ model which includes contributions from electronic states with $S$ and $P$ symmetries to model, e.g., HgTe quantum wells.  Here, we follow a derivation by Lunde and Platero \cite{Lunde_2013}, who estimated the hyperfine interaction for that model by averaging over nuclear spin locations within the edge state.  Transforming their expression into real space, we have
\begin{equation}
H_n^\text{HF}=F(x-\tfrac{x_{n-1}+x_{n}}{2}) \left[ A^z\sigma_z\tau_z + A^\perp \left(\sigma_-\tau_++\sigma_+\tau_- \right)\right],\label{hf-BHZ}
\end{equation}
where $A^z$ and $A^\perp$ control the anisotropic coupling, $\tau_i$ are nuclear spin Pauli matrices, and $F$ is a spatial form factor.  We will later see that our results are mostly independent of the form factor.  Each $H_n^\text{HF}$ violates the usual electronic time-reversal symmetry, but preserves the generalized time-reversal invariance (GTRI) that flips the electronic momentum as well as \emph{both} the electronic and nuclear spins.  The nuclear spins are assumed to have $S=1/2$, and be sparse enough that at most one $H^\text{HF}_n(x)$ interaction is nonzero at any $x$, vanishing outside some interval $[x_{n-1},x_n]$.  We have previously \cite{Vezvaee_2018} considered this interaction on a ring, for small numbers of nuclear spins $N$.

\begin{figure}
\includegraphics{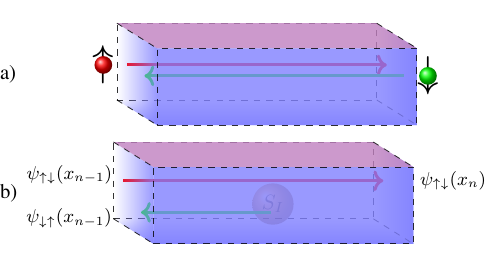}
\caption{(a) Topological insulator without TR-breaking impurities, and concomitant protected edge currents.  (b) A nuclear spin allows a propagating electron to backscatter.  The rectangular box represents the region over which the electron interacts with the $n$th nuclear spin.  This region is assumed to be finite, but the interaction strength can vary arbitrarily over this region.
\label{figure-cartoon}}
\end{figure}

\section{Single Nuclear Spin\label{sec:single}}
For the moment, consider only the $n$th nuclear spin, where the wavefunction with electron at $x$ and electronic and nuclear $z$ spin projections $m_e$ and $m_I$ is $\psi_{m_e,m_I}(x)$.  Because \eqnref{full-H} is first-order, its eigenstates are completely determined by fixing the value of the wavefunction at any single point.  The eigenstates must be continuous across the boundaries of the interaction region.  If we consider an eigenstate corresponding to an electron incoming from the left, then there are two possibilities: either the electron and nuclear spins are parallel, in which case no backscattering can occur, or they are antiparallel, in which case the electron has a nonzero probability to be reflected, as shown in \figref{figure-cartoon}.  In the former case, the parallel-spin wavefunction components at $x_{n-1}$ and $x_n$ (the left and right sides of the interaction region) must be equal up to a phase, which we denote as $p$:
\begin{equation}
\frac{\psi_{\uparrow\uparrow}(x_{n})}{\psi_{\uparrow\uparrow}(x_{n-1})}=\frac{\psi_{\downarrow\downarrow}(x_{n})}{\psi_{\downarrow\downarrow}(x_{n-1})}=p.
\end{equation}
In the case of antiparallel spins, the corresponding wavefunction components are related by reflection and transmission amplitudes:
\begin{equation}
\frac{\psi_{\uparrow\downarrow}(x_{n})}{\psi_{\uparrow\downarrow}(x_{n-1})}=\frac{\psi_{\downarrow\uparrow}(x_{n-1})}{\psi_{\downarrow\uparrow}(x_{n})}=t,
\end{equation}
\begin{equation}
\frac{\psi_{\downarrow\uparrow}(x_{n-1})}{\psi_{\uparrow\downarrow}(x_{n-1})}=r_{\hookleftarrow}\quad\text{and}\quad \frac{\psi_{\uparrow\downarrow}(x_{n})}{\psi_{\downarrow\uparrow}(x_{n})}=r_{\hookrightarrow},
\end{equation}
Here, we have allowed for the possibility that the scattering amplitudes can differ depending on whether the electron is incoming from the left or from the right.  In particular, $r_{\hookleftarrow}$ ($r_{\hookrightarrow}$) is the reflection amplitude for an electron incoming from the left (right).  Notice that we have taken the transmission amplitude $t$ and the ``passing'' amplitude $p$ to be the same regardless of where the incoming wave comes from.  As we show in Appendix \ref{appendix:symmetry}, GTRI imposes a left-right symmetry on these amplitudes.  In addition to this symmetry, GTRI also imposes two more constraints on the scattering amplitudes:

\begin{equation}
|r_{\hookleftarrow}|^2=|r_{\hookrightarrow}|^2=1-|t|^2=:|r|^2,\label{mag-r}
\end{equation}
\begin{equation}
 r_{\hookleftarrow}r_{\hookrightarrow} = - \frac{t^2}{|t|^2} |r|^2.\label{phase-r}
\end{equation}

The second equation (see \eqnref{eqn:rt-relation}) ``forgets'' the phases accumulated at a site after being flipped twice---this key observation ultimately allows for the simplification of the problem.  It is important to note that these constraints hold regardless of the shape of the interaction profile $F$, even if it is spatially asymmetric.  Explicit expressions for $r_{\hookleftarrow}$, $r_{\hookrightarrow}$, $t$, and $p$ for a chosen $F$ can of course be obtained by solving \eqnref{full-H} inside and outside the interaction region and by imposing wavefunction continuity across the boundaries of this region.  This is done for the case of a square profile in Appendix \ref{appendix:barrier}.  Remarkably, as we will show, it is possible to obtain an expression for the total reflection amplitude in the case of many nuclear spins solely in terms of the scattering amplitudes $r_{n,\hookleftarrow}$, $r_{n,\hookrightarrow}$, $t_n$, $p_n$ for the individual nuclei.

Before we show how the single-nucleus scattering data $r$, $t$, and $p$ can be used to construct scattering amplitudes for arbitrarily many nuclear spins, we first estimate the physical values of these parameters for the case of HgTe.  For simplicity, we consider a square interaction profile, $F(\delta)=\Theta(|\delta|-w/2)/w$, where $\Theta$ is the Heaviside function---i.e., a square barrier of width $w$ and unit total area centered in the interaction region.  Defining $L_n=x_n-x_{n-1}$, we obtain
\begin{equation}
p_n=e^{iEL_n/\hbar v_0} e^{-iA^z/\hbar v_0},
\end{equation}
as shown in Appendix \ref{appendix:barrier}.  The calculation of $r$ and $t$ amounts to boundary matching on the two-dimensional subspace with zero total angular momentum and (squared) linear momentum $(\hbar k)^2$.  The result is
\begin{multline}
t= e^{iE(L_n-w_n)/\hbar v_0} \mathrm{exp}\left[\cot^2(\theta)/\sin^2(|A^\perp/\hbar v_0|\cot\theta)\right]^{-1/2}\\
\times \left[-i\arctan\frac{-\sin(|A^\perp/\hbar v_0|\cot\theta)}{\cos\theta\cos(|A^\perp/\hbar v_0|\cot\theta)}\right],
\end{multline}
where $\csc\theta=(Ew+A^z)/|A^\perp|$.  Reflection coefficients follow from \eqnsref{mag-r}{phase-r}, but the phases depend on an arbitrary partitioning of the line (see Appendix \ref{appendix:barrier}).  The overall factor $e^{iEL_n/\hbar v_0}$ common to all of the parameters indicates that the spacing of the nuclear moments changes only the phases of the parameters, which does not materially affect the behavior of the system (see Appendix \ref{appendix:spin-migration}).

Lunde et.\ al \cite{Lunde_2013} estimate $A^\perp$ and $A^z$ for the spinful isotopes of mercury and tellurium, finding that the hyperfine coupling of tellurium is an order of magnitude stronger than that of mercury.  We therefore neglect mercury's hyperfine coupling, and focus only on tellurium's contribution, for which\footnote{Platero finds both $P$ and $S$ contributions to the hyperfine coupling, but our scattering results depend on the barrier area, with units energy-length (and our $A^z$ and $A^\perp$ consequently include a length factor).  The contact interaction is therefore neglected, and only the $P$ portion (approximately 10\%) contributes.  We assume the length scale $\sim \mathrm{\AA}$.}
\begin{equation}
|r_n|^2 = (A^\perp /\hbar v_0)^2\sim 10^{-15}.
\end{equation}
In the following, our results remain exact even for $|r|^2\sim 1$, though our focus will be on the physically relevant regime of small $|r|^2$.

\begin{figure}
\includegraphics{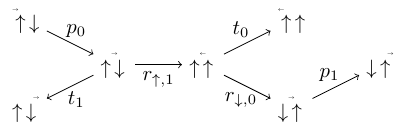}
\caption{A graphical representation of the boundary matching procedure determining the eigenstate $\ket{\Psi_{\uparrow,\downarrow;\rightarrow}^{(E)}}$.  The compact notation $\stackrel{.\hspace{0.5em}}{\uparrow}\stackrel{\hspace{0.5em}}{\downarrow}$ describes a state with first nuclear spin up, and second nuclear spin down, and the small arrow indicates the electron location (its spin can be inferred by conservation of angular momentum).  Edges connect amplitudes which are related by passing, transmission, or reflection.  Although this graph is a tree (i.e., there is at most one path between nodes), generically there will be multiple paths connecting states.  A simpler example of such boundary matching is worked out in Appendix \ref{appendix:naive-direct}.  \label{figure-graph}}
\end{figure}

\section{Multiple Nuclei\label{sec:multi}}
In the multi-nuclear case ($N>1$), the process of boundary matching remains straightforward, but the computation time scales exponentially with $N$ (see Appendix \ref{appendix:naive-direct}).  Let $\ket{\Psi_{\mathbf{m};\rightarrow}^{(E)}}$ be the energy $E$ eigenstate with electron incoming from the left, and with an ``initial'' nuclear spin configuration $\mathbf{m}=\left\{m_{n}\right\}$.  In particular, its components
\begin{equation}
\mathcal{Z}_{\mathbf{m}\mathbf{m}',j} = \braket{x_j; m_e'(\mathbf{m},\mathbf{m}');\mathbf{m}'}{\Psi_{\mathbf{m};\rightarrow}^{(E)}}\label{Z-definition}
\end{equation}
completely characterize this eigenstate.  $\ket{x_j; m_e'(\mathbf{m},\mathbf{m}');\mathbf{m}'}$ is the state with electron at $x_j$, nuclear spin configuration $\left\{m_{n}'\right\}$, and (unique) electronic spin $m_e'(\mathbf{m},\mathbf{m}')$ allowed by conservation of angular momentum.  \figref{figure-graph} shows a graphical representation of a single scattering eigenstate constructed from the standard boundary matching process for the case of $N=2$ nuclear spins.  The nonzero amplitudes for this state are vertices connected by edges representing passing, transmission, and reflection.  The graph shows how each amplitude is successively calculated by multiplying single-nucleus scattering amplitudes $r$, $p$, $t$ together along paths connecting each possible final spin configuration to the initial spin configuration.  More complex examples for $N=3$ nuclear spins are worked out in detail in Appendix \ref{appendix:naive-direct}.  With this approach, it is straightforward to show that each amplitude is obtained by summing over all such paths:
\begin{equation}
\mathcal{Z}_{\mathbf{m}\mathbf{m}',j'} = \sum_{P} \prod_n r_{\hookrightarrow,n}^{N_{r_{\hookrightarrow,n}}[P]}  r_{\hookleftarrow,n}^{N_{r_{\hookleftarrow,n}}[P]}p_n^{N_{p_n}[P]}t_n^{N_{t_n}[P]},\label{Z-pathsum}
\end{equation}
where $N_{r_{\hookrightarrow,n}}[P]$ is the number of reflections from the right at site $n$ along path $P$, and $N_{r_{\hookleftarrow,n}}[P]$, $N_{p_n}[P]$, and $N_{t_n}[P]$ are defined similarly (see Appendix \ref{appendix:naive-direct}).  Surprisingly, this expression can be drastically simplified to the form,
\begin{multline}
\mathcal{Z}_{\mathbf{m}\mathbf{m}',j'} = \mathcal{Z}_{0;\mathbf{m},\mathbf{m}'} \\
\times \prod_n \frac{1}{2}\left[(1+z_n)^{N^0_n} +(-1)^{|\delta J_n|}(1-z_n)^{N^0_n}\right],\label{Z-calculated}
\end{multline}
as shown in Appendix \ref{appendix:spin-migration}.  Here, $z_n=i|r_n/t_n|$, and $\delta J_n=(m_{n}'-m_{n})/2$ is the change in the $n$th nuclear spin.  The quantity $N^0_n=N_{t,n}+N_{r_{\hookrightarrow,n}}+N_{r_{\hookleftarrow,n}}$ (see \eqnref{def:surplus}) is found to be path independent, as is the overall prefactor $\mathcal{Z}_{\mathbf{m}\mathbf{m}',j'}$ (see \eqnref{eqn:Z-Z_0-pathcounting}).   All of these are easily calculated, so that the time needed to evaluate \eqnref{Z-calculated} grows linearly with the number of nuclear spins $N$.  Other approaches (including direct Hermitian diagonalization) require exponential computation time.  This dramatic speedup allows us to study the dynamics for realistic system sizes.

Before discussing how to compute \eqnref{Z-calculated}, there are some observations that can be made immediately from this analytical expression.  The magnitude of each outgoing amplitude $\mathcal{Z}_{\mathbf{m}\mathbf{m}';j'}$ is independent of the phases of $t$ (and $r$).  Moreover, notice that moving any of the impurity interactions only serves to change the phases of each outgoing amplitude.  Remarkably, this means that the spacing between impurities does not affect the behavior when a single electron is passed through the system (in fact, this generalizes to multiple passes of electrons).  Furthermore, $\mathcal{Z}_{\mathbf{m}\mathbf{m}';j'}$ is a product of smooth functions, each depending on only one $z_n$.  Notice that phase changes with $N\to\infty$ are still possible, but only when the net effect of a large number of $z_n$ contribute coherently.  Compare this with Anderson localization, where disorder in individual terms (each of which contribute incoherently) can lead to a phase transition.  Here, of course, there are several differences from the original Anderson model that make it a priori unclear whether localization should be expected to occur.  The first is that we are really considering a many-body system since we retain the full quantum mechanical degrees of freedom of the nuclear spin lattice.  Secondly, though a linearly-dispersing system may fail to transmit an electron from one side to the other, Klein tunneling will prevent any truly localized state.

\begin{figure}
\includegraphics{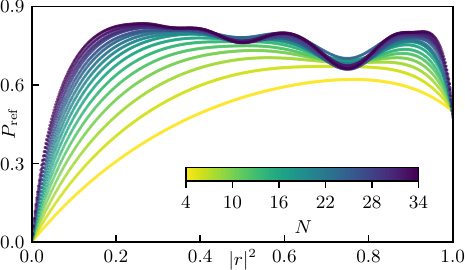}
\caption{Plots of the probability of reflection $P_\mathrm{ref}$ for various numbers of nuclear spins, $N$.  An average is performed over randomly chosen initial spin configurations with no net initial polarization (i.e., equal number of up and down spins).  For each value of $N$, we average over either $1024$ realizations, or a complete survey of the sample space, whichever is smaller.  \label{figure-zerofield-P_l}}
\end{figure}

To demonstrate the power of this approach, we consider the case of uniformly distributed nuclear spins, without disorder or initial polarization.  The single-spin interactions are completely characterized by the common magnitude $|r|^2=|r_n|^2$ (seen from \eqnsref{mag-r}{Z-calculated}).  The overall probability of reflection, $P_\mathrm{ref}$, i.e., the probability that an electron injected at the left side of the TI will exit on the left, is obtained by summing the squares of the amplitudes in \eqnref{Z-calculated} that correspond to an electron departing the nuclear spin lattice on the left, and is plotted in \figref{figure-zerofield-P_l} for an ensemble of initial spin configurations, each with equal numbers of up and down spins.

Notice that, when $|r|^2=0$, the material is in the perfect backscatter-free conducting limit.  At $|r|^2=1$, however, the system behaves pseudo-classically: each initial spin configuration has exactly one outgoing spin configuration.  In fact, if the first nucleus and incoming electron are spin up, transmission through the lattice is guaranteed.  Every down spin met by the electron will result in two immediate backscatters.  Thus, if the first nuclear spin is up, then the electron is ultimately forced to move to the right, eventually transmitting through the entire lattice.  If the first nuclear spin is down, the electron is backscattered by the lattice.  In the zero net nuclear polarization case, this translates to an overall backscatter probability of $1/2$ in this ``perfect scatterer'' limit (see \figref{figure-zerofield-P_l}).

\begin{figure}
\includegraphics{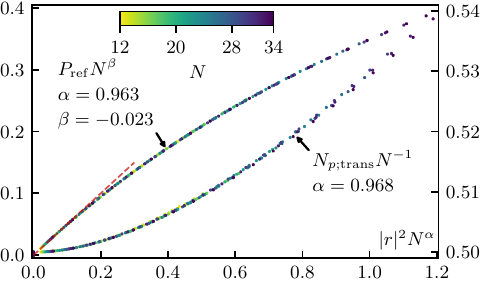}
\caption{Scaling of the overall reflection probability, and number of passes for systems with $12$ though $34$ nuclear spins, and no overall initial magnetization (1024 realizations).  Data is restricted to $|r|^2< 0.04$, to focus on the effects of the zero-$r$ phase transition.  \label{figure-at0-zerofield}}
\end{figure}

Given this comment, there must be a phase transition from the perfect conductor at $|r|^2=0$ to the ``perfect scatterer'' at $|r|^2=1$.  Indeed, \figref{figure-zerofield-P_l} shows many interesting features, including several local maxima in overall reflection.  These features perhaps indicate additional phases beyond the two identified above.  Though interesting, these ``large $|r|^2$'' features are not studied in detail here since the physical system of interest, topological insulators interacting with either nuclear or atomic magnetic moments, will have $|r|^2 \ll 0.1$, far below these other features.

\section{Scaling\label{sec:scaling}}
Next, we show that the total reflection probability and other scattering information exhibit universal scaling as the number of nuclear spins $N$ is increased.  This result is both surprising and crucial for making predictions for real physical systems such as topological insulators, where $N\gg1$.  If $N_\downarrow$ ($N_\uparrow$) is the number of spins initially down (up), the change in nuclear polarization per injected electron can be bounded above by noticing that each down spin acts as a scattering source:
\begin{equation}
 \frac{\partial N_\uparrow}{\partial j} \leq |r|^2\sum_{n=0}^{N_\downarrow-1} (1-|r|^2)^{n} = 1-(1-|r|^2)^{N_\downarrow}.
\end{equation}
In the extreme $|r|^2\to0$ limit (i.e., semi-classical weak scattering) the equality becomes exact (because subsequent reflections are higher order in $|r|^2$), and we obtain
\begin{equation}
 \frac{\partial N_\uparrow}{\partial j} = |r|^2 N_\downarrow.\label{scaling-diffeq}
\end{equation}
\figref{figure-at0-zerofield} shows the agreement with this limit, with a line of slope $1/2$ at $|r|^2=0$ to guide the eye.  While the behavior for $|r|^2N_\downarrow\ll1$ is easily found by this argument, when $|r|^2N_\downarrow\sim 1$, a reflection is expected to occur before the electron crosses the lattice.  In fact, many reflections are likely to occur in this limit, and the quantum mechanical phases become very important for the calculation.  Notice that the universal scaling in \figref{figure-at0-zerofield} persists far past the trivial linear limit, and moreover up to at least $|r|^2N_\downarrow\sim 1$, well inside this quantum regime.

Another measure of the system's response to the incident electron is $N_{p,n}$, the number of passes at site $n$.  It is path independent, like $N^0_n$, and corresponds precisely to the distance up spins have migrated in the spin-momentum locked direction of motion (see \eqnref{def:d_S} for a precise discussion of this quantity).  This is most easily seen for a totally up-spin polarized system with $N$ sites: the electron itself ``passes'' $N$ times through this system (i.e., there are $N$ spin-parallel approaches).  The quantity $N_{p;\mathrm{trans}}$ in \figref{figure-at0-zerofield} is the sum of $N_{p,n}$ over all sites, \emph{conditioned} on there being an overall transmission event.  The scaling seen is consistent with the $N_\uparrow|r|^2\ll1$ limit.

\begin{figure}
\includegraphics{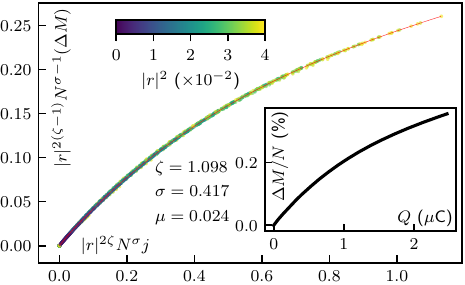}
\caption{Scaling plot of change in magnetization $\Delta M$ as a function of the (scaled) number of injected electrons $j$.  The scaling collapses for $N=6,8,10,12,14$, using all states with $M=0$.  The red line is a numerical fit of the scaling function $\mathfrak{g}$ to a quartic spline (it is, in fact, poorly fit by a simple exponential).  Data is restricted to $|r|^2<0.04$, to focus on the effects of the zero-$r$ phase transition.  (inset) Predicted nuclear polarization as a function of injected charge (in micro-Coulombs) for system with realistic scattering $|r|^2=10^{-15}$ and $N=10^8$ spinful nuclei.  We set $\mu=\delta=0$ to maintain numerical stability.\label{figure-all-current}}
\end{figure}

Both of these measures suggest scaling behavior near $|r|^2=0$, which extends to another property: the dynamic nuclear spin polarization that results when current is injected into such a system.  The calculation is also done numerically exactly.  Up to $12$ electrons are injected into $N\leq 16$ nuclear spins, keeping all quantum mechanical phases, for systems without disorder or initial polarization, and assuming the edge current is small enough that each electron propagates through the nuclear spin lattice independently.  For $|r|^2\ll 1$, we again find scaling: the total nuclear polarization is a function of only
\begin{equation}
M = |r|^{-2\beta}N^{-\gamma} \mathfrak{g}(|r|^{2\zeta} N^\sigma j(1+\mu j+\delta j^2)),\label{scaling-current}
\end{equation}
with numerically identified scaling function $\mathfrak{g}$ and fit parameters $\beta,\zeta,\sigma,\mu,\delta$ (see \figref{figure-all-current}).

To understand why such scaling might occur, and to constrain the collapse, differentiate \eqnref{scaling-current}, combine it with \eqnref{scaling-diffeq}.  Near $|r|^2=0$,
\begin{equation}
\frac{N}{2} |r|^2 = |r|^{2(\zeta-\beta)} N^{\sigma-\gamma} \mathfrak{g}'(0),
\end{equation}
requiring $\zeta-\beta=1$, $\sigma-\gamma= 1$, and $\mathfrak{g}'(0)= 1/2$, enforced on the collapse.\footnote{Assuming $\zeta,\sigma>0$, which are indeed numerically found to be the case}

Three different kinds of scaling are responsible for the collapse: individual scaling, ensemble scaling, and $N$ scaling.  The $M_{\mathbf{m}}(j)$ polarization functions scale individually: for a given initial spin configuration $\mathbf{m}$, the expected polarization obeys \eqnref{scaling-current}, though the $N$ dependence is trivial.  In principle, a generic dependence on the number of injected electrons $j$, not a simple linear scaling.  Amazingly, the collapse is very nearly a linear dependence on $j$, with $|\mu|< 3\times 10^{-2}$ and the higher term $|\delta| < 3\times 10^{-3}$.  Given this individual scaling, the \emph{ensemble} averages automatically scale, though in principle such ensemble scaling could exist without individual scaling.  Finally, ensembles with different $N$ obey \eqnref{scaling-current}, with nontrivial $N$ dependence.

\eqnref{scaling-current} enables precise microscopic predictions for electronic edge transport through realistic systems, at a specified energy.  The edge states of a HgTe quantum well of thickness $10\,\mathrm{nm}$ and side length $100 \mathrm{\mu m}$ interact with approximately $N\sim10^{8}$ spinful nuclei (for an edge state penetration depth of $\sim50$ nm \cite{Lunde_2013}).  The power of this scaling is demonstrated in the inset of \figref{figure-all-current}, where we show the resulting prediction for the nuclear polarization in this full-scale quantum well with macroscopic injected currents.  To detect these effects experimentally, NMR studies \cite{Mukhopadhyay_2015} could be performed and the signals compared before and after current passage.  Because of the limited resolution of traditional NMR, novel techniques with nanoscale resolution may be preferable.  For example, an individual NV center in diamond has been recently used for sensing of proton nuclear magnetic resonance in an organic sample \cite{Rugar_2014}.  In addition, the low-entropy configurations created by the nuclear polarization might allow energy to be stored and extracted via Landauer's principle \cite{Bozkurt_2018}.

\section{Conclusion\label{sec:conclusion}}
In conclusion, we have solved the electron-nuclear scattering problem in 2D TI edges for macroscopic numbers of nuclear spins by exploiting the symmetries of the problem, dramatically speeding up the numerical computation, and by leveraging a surprising universal scaling behavior.  Our solution reveals that modest edge currents can generate significant nuclear spin polarization that should be detectable in current TI experiments.

\section*{Acknowledgements}
This research was supported by the NSF (Grant No. 1741656).

\bibliography{references}

\onecolumngrid
\appendix
\section{Single Nuclear Spin\label{appendix:barrier}} 
\begin{figure}
\includegraphics{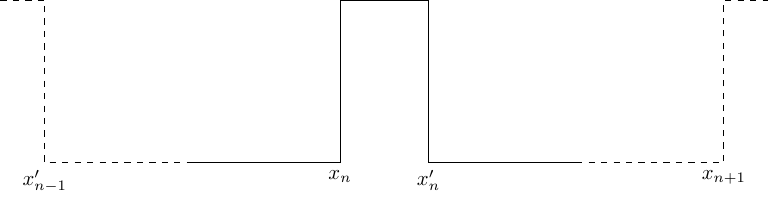}
\caption{Alternative parameterization of eigenstates of \eqnref{hf-BHZ}.  The (dashed) left and right square potentials correspond to interactions with other nuclei (and serve only as a guide to the eye).  \label{fig:primeconvention}}
\end{figure}

Here, we consider a single spin-$1/2$ nucleus, interacting with the electron spin in the region $[x_n,x_n']$, with square interaction profile (as in \eqnref{hf-BHZ}).
To allow for spacing between nuclear spins, we also consider the point $x_{n+1}>x'_n$, and consider the total length $L_n=x_{n+1}-x_n$.  The width of the square well is $w_n= \quad x'_n - x_{n}$.  For a single nuclear spin, $n=1$, $x_n=0$, $x'_n=w_0$, $x_2=L_1$.  The index (and $L_1$) are of course pointless.  However, for multiple nuclei this will establish a convention.  $L_n$ will serve as the total distance between the left sides of each square well interacting region (see \figref{fig:primeconvention}).  Inside the interaction region, in the basis $\ket{\uparrow\uparrow},\ket{\uparrow\downarrow},\ket{\downarrow\uparrow},\ket{\downarrow\downarrow}$
(the first spin corresponds to the electron, the second to the nuclear spin), the Hamiltonian is
\begin{equation}
H_q = \begin{bmatrix} A^z/w_n+q \\ & B_q \\ && A^z/w_n - q\end{bmatrix}.\label{eqn:Bq-hamiltonian}
\end{equation}
We have set $\hbar v_0=1$ in the above for simplicity.  $B_q$ is the 2 by 2 matrix
\begin{equation}
B_q = \begin{bmatrix} q-A^z/w_n& \overline{A^\perp}/w_n \\ A^\perp/w_n & -q-A^z/w_n \end{bmatrix} = -(A^z/w_n) \mathbf{1} + d\lB \sigma_z \cos\theta + \sigma_x \cos(\phi) \sin\theta + \sigma_y \sin(\phi) \sin\theta\rB,
\end{equation}
where
\begin{equation}
d^2 = |A^\perp/w_n|^2 + q^2.
\end{equation}
Outside $[x_n,x_n']$, $A^z=A^\perp=0$ and possibly a different $q$.  The phase $\phi$ defined by $A^\perp=|A^\perp|e^{i\phi}$ can be set to zero
\begin{equation}
\begin{bmatrix} \cos\theta\\ \sin\theta \end{bmatrix} = \frac{1}{d}\begin{bmatrix} q \\\ |A^\perp/w_n| \end{bmatrix}.
\end{equation}

The eigenvectors of $B_q$ correspond to the eigenstates of the Hamiltonian that couple the electronic and nuclear spins, and are
\begin{equation}
\begin{bmatrix} \cos\theta/2\\ e^{i\phi} \sin\theta/2 \end{bmatrix}\quad\text{and}\quad \begin{bmatrix} -\sin\theta/2\\ e^{i\phi}\cos\theta/2 \end{bmatrix},\label{squarewell-eigenvecs}
\end{equation}
with eigenvalues
\begin{equation}
 E = -A^z/w_n\pm d\quad\text{where}\quad |q| = \sqrt{(E+A^z/w_n)^2-|A^\perp/w_n|^2},\label{squarewell-eigenenergy}
\end{equation}
where
\begin{equation}
\cot\theta = \frac{q}{|A^\perp/w_n|} = \pm \sqrt{(E+A^z/w_n)^2/|A^\perp/w_n|^2-1}.
\end{equation}

Outside $[x_n,x_n']$, any electronic spin-up eigenstate's wavefunction must be proportional to $e^{iEx}$ (respectively $e^{-iEx}$ for electronic spin down).  Routine decomposition of $\psi(x_n)$ and $\psi(x_n')$ into the eigenvectors of the interacting and noninteracting Hamiltonian \eqnref{eqn:Bq-hamiltonian} (with appropriate $q$ to give the correct eigenenergy), along with continuity of the wavefunction inside and outside $[x_n,x_n']$, lead to conditions on the wavefunction.  Solving for the form of the wavefunction inside $[x_n,x_n']$ leads to the conditions.  Next, we outline this boundary matching process.

The first and fourth component decouple from each other and all other eigenstates:
\begin{equation}
e^{-iE(L_n-w_n)}\psi_{\uparrow\uparrow}(x_{n+1}) =
\psi_{\uparrow\uparrow}(x'_n) = e^{i(E-A^z/w_n)w_n}\psi_{\uparrow\uparrow}(x_n)
\end{equation}
and
\begin{equation}
e^{iE(L_n-w_n)}\psi_{\downarrow\downarrow}(x_{n+1}) =
\psi_{\downarrow\downarrow}(x'_n) =
e^{i(A^z/w_n-E)w_n}\psi_{\downarrow\downarrow}(x_n).\label{eqn:xn-bc}
\end{equation}

The decomposition of the second and third components into the eigenstates \eqnref{squarewell-eigenvecs} does not decouple:
\begin{equation}
\begin{bmatrix}\psi_{\uparrow\downarrow}(x_n)\\\psi_{\downarrow\uparrow}(x_n)\end{bmatrix}
=
\begin{bmatrix}\cos\theta_n/2\\ \sin\theta_n/2\end{bmatrix}
c_n^++
\begin{bmatrix}\sin\theta_n/2\\ \cos\theta_n/2\end{bmatrix}
c_n^-.
\end{equation}
Notice the lack of minus sign in the eigenvector multiplying $c_n^-$.  The $c_n^+$ and $c_n^-$ are coefficients of eigenstates of $B_{q'}$ and $B_{-q'}$, respectively, where $q'$ is positive and given by \eqnref{squarewell-eigenenergy} with the same energy $E$.  These eigenstates are therefore not generically orthogonal, though they are linearly independent except for forbidden energies (cf., \cite{Vezvaee_2018}).  At $x'_n$,
\begin{equation}
\begin{bmatrix}e^{-iE(L_n-w_n)}\psi_{\uparrow\downarrow}(x_{n+1})\\e^{iE(L_n-w_n)}\psi_{\downarrow\uparrow}(x_{n+1})\end{bmatrix}
=
\begin{bmatrix}\psi_{\uparrow\downarrow}(x'_n)\\\psi_{\downarrow\uparrow}(x'_n)\end{bmatrix}
=
\begin{bmatrix}\cos\theta_n/2\\ \sin\theta_n/2\end{bmatrix}
c_n^+e^{iq'w_n}+
\begin{bmatrix}\sin\theta_n/2\\ \cos\theta_n/2\end{bmatrix}
c_n^-e^{-iq'w_n}.
\end{equation}
We capture the left-right symmetry of the problem by using the above linear equations to express, in terms of $c_n^\pm$, the \emph{incoming} amplitudes,
\begin{equation}
\begin{bmatrix} \psi_{\uparrow\downarrow}(x_n) \\ \psi_{\downarrow\uparrow}(x_{n+1}) \end{bmatrix}
=
\begin{bmatrix}
 \cos\theta_n/2&\sin\theta_n/2\\
e^{i(-E(L_n-w_n)+q'w_n)}\sin\theta_n/2&e^{i(-E(L_n-w_n)-q'w_n)}\cos\theta_n/2\\
\end{bmatrix}
\begin{bmatrix} c_n^+ \\ c_n^-\end{bmatrix},\label{bc-incoming}
\end{equation}
and \emph{outgoing} amplitudes,
\begin{equation}
\begin{bmatrix} \psi_{\uparrow\downarrow}(x_{n+1}) \\ \psi_{\downarrow\uparrow}(x_{n}) \end{bmatrix}
=
\begin{bmatrix}
e^{i(E(L_n-w_n)+q'w_n)}\cos\theta_n/2&e^{i(E(L_n-w_n)-q'w_n)}\sin\theta_n/2\\
 \sin\theta_n/2&\cos\theta_n/2
\end{bmatrix}
\begin{bmatrix} c_n^+ \\ c_n^-\end{bmatrix}.\label{bc-outgoing}
\end{equation}
Invert \eqnref{bc-incoming} to get
\begin{equation}
\begin{bmatrix} c_n^+ \\ c_n^-\end{bmatrix}
=\Omega
\begin{bmatrix}
e^{-iq'w_n}\cos\theta_n/2&-e^{iE(L_n-w_n)}\sin\theta_n/2\\
-e^{iq'w_n}\sin\theta_n/2&e^{iE(L_n-w_n)}\cos\theta_n/2\label{bc-incoming-inv}
\end{bmatrix}
\begin{bmatrix}\psi_{\uparrow\downarrow}(x_n)\\\psi_{\downarrow\uparrow}(x_{n+1})\end{bmatrix},
\end{equation}
where
\begin{equation}
\Omega^{-1}=
e^{-iq'w_n}\cos^2\theta_n/2-e^{iq'w_n}\sin^2\theta_n/2
= \sqrt{1-\cos^2q'w_n\sin^2\theta_n}
 \exp\lP i\arctan \frac{\sin q'w_n}{\cos q'w_n \cos \theta_n}\rP .
\end{equation}
Substitute \eqnref{bc-incoming-inv} into \eqnref{bc-outgoing} to get
\begin{multline}
\begin{bmatrix}
\psi_{\uparrow\downarrow}(x_{n+1}) \\ \psi_{\downarrow\uparrow}(x_{n})
\end{bmatrix}
=\\
\Omega\begin{bmatrix}
e^{iE(L_n-w_n)}\lB \cos^2\theta_n/2 - \sin^2 \theta_n/2 \rB &
-2i e^{2iE(L_n-w_n)}\sin\lP q'w_n \rP \cos\theta_n/2\sin\theta_n/2 \\
-2i \sin\lP q'w_n \rP \cos\theta_n/2\sin\theta_n/2 &
e^{iE(L_n-w_n)}\lB \cos^2\theta_n/2 - \sin^2 \theta_n/2 \rB
\end{bmatrix}
\begin{bmatrix}
\psi_{\uparrow\downarrow}(x_{n+1}) \\ \psi_{\downarrow\uparrow}(x_{n})
\end{bmatrix}.
\end{multline}

Collecting the results for all four components, we have related all outgoing amplitudes to the incoming amplitudes,
\begin{equation}
\begin{bmatrix} \psi_{\uparrow\uparrow}(x_{n+1}) \\ \psi_{\uparrow\downarrow}(x_{n+1}) \\ \psi_{\downarrow\uparrow}(x_{n}) \\ \psi_{\downarrow\downarrow}(x_{n}) \end{bmatrix} =
\begin{bmatrix}p_n&&&\\&t_n&r_{n,\hookrightarrow}&\\&r_{n,\hookleftarrow}&t_n&\\&&&p_n\end{bmatrix}
\begin{bmatrix} \psi_{\uparrow\uparrow}(x_{n}) \\ \psi_{\uparrow\downarrow}(x_{n}) \\ \psi_{\downarrow\uparrow}(x_{n+1}) \\ \psi_{\downarrow\downarrow}(x_{n+1}) \end{bmatrix},\label{eqnscatmatrix}
\end{equation}
where noting that $q'w_n=|A^\perp_n|\cot\theta$, after simplification,
\begin{align}
r_{n,\hookrightarrow} &= ie^{2iE(L_n-w_n)}
\lB1+\frac{\cot^2\theta}{\sin^2(|A^\perp_n|\cot\theta)}\rB^{-1/2} \exp\lB-i\arctan\frac{-\sin(|A^\perp_n|\cot\theta)}{\cos\theta\cos(|A^\perp_n|\cot\theta)}\rB,\nonumber\\
r_{n,\hookleftarrow} &= i
\lB1+\frac{\cot^2\theta}{\sin^2(|A^\perp_n|\cot\theta)}\rB^{-1/2} \exp\lB-i\arctan\frac{-\sin(|A^\perp_n|\cot\theta)}{\cos\theta\cos(|A^\perp_n|\cot\theta)}\rB,\nonumber\\
t_n &= e^{iE(L_n-w_n)}
\lB\frac{\cot^2\theta}{\sin^2(|A^\perp_n|\cot\theta)}\rB^{-1/2} \exp\lB-i\arctan\frac{-\sin(|A^\perp_n|\cot\theta)}{\cos\theta\cos(|A^\perp_n|\cot\theta)}\rB,\nonumber\\
p_n &=e^{i(EL_n-A^z_n)}.\label{eqn:full-rtp}
\end{align}

$p$ is named for ``passing,'' as it corresponds to the solutions in which conservation of angular momentum and spin-momentum locking forbid electronic backscatter.  The $r$ and $t$ coefficients are named for reflection and transmission, respectively.  The $r_{\hookrightarrow}$ and $r_{\hookleftarrow}$ refer to reflection outgoing to the right and left, respectively.  The two physical scenarios of left-incoming electron with parallel and antiparallel nuclear spins can be described by particular eigenstates satisfying \eqnref{eqnscatmatrix}, respectively $\ket{\phi_{\uparrow\uparrow}}$ and $\ket{\phi_{\rightarrow}}$:
\begin{equation}
\braket{x_n}{\phi_{\uparrow\uparrow}} = \begin{bmatrix} 1\\ 0\\0\\0 \end{bmatrix}
,\quad\text{and}\quad
\braket{x_{n+1}}{\phi_{\uparrow\uparrow}} = \begin{bmatrix} p_n\\ 0\\0\\0 \end{bmatrix},
\end{equation}

\begin{equation}
\braket{x_n}{\phi_{\rightarrow}} = \begin{bmatrix} 0\\ 1\\r_{n,\hookleftarrow}\\0 \end{bmatrix}
,\quad\text{and}\quad
\braket{x_{n+1}}{\phi_{\rightarrow}} = \begin{bmatrix} 0\\ t_n\\0\\0 \end{bmatrix}.
\end{equation}

Similarly, the two right-incoming scenarios with parallel and antiparallel nuclear spin have respective wavefunctions $\ket{\phi_{\downarrow\downarrow}}$ and $\ket{\phi_{\leftarrow}}$:

\begin{equation}
\braket{x_n'}{\phi_{\downarrow\downarrow}} = \begin{bmatrix} 0\\ 0\\0\\1 \end{bmatrix},
\quad\text{and}\quad
\braket{x_{n-1}'}{\phi_{\downarrow\downarrow}} = \begin{bmatrix} 0\\ 0\\0\\p_n \end{bmatrix},
\end{equation}

\begin{equation}
\braket{x_{n+1}}{\phi_{\leftarrow}} =
 \begin{bmatrix} 0\\ r_{n,\hookrightarrow}\\1\\0 \end{bmatrix},
\quad\text{and}\quad
\braket{x_{n-1}'}{\phi_{\leftarrow}} = \begin{bmatrix} 0\\ 0\\t_n\\0 \end{bmatrix}.
\end{equation}

These four solutions fully characterize the solution space.  In this picture, it is clear that simply evaluating the wavefunction at a different location, i.e. placing the square well symmetrically, can remove the phase difference between the left and right reflection coefficients; they are unequal only because the square well potential is not centered in $[x_n,x_{n+1}]$.  Setting $x'_n=x_{n+1}$ removes the asymetry, and $r_n=r_{n,\hookrightarrow}=r_{n,\hookleftarrow}$.  In that case, a relative phase factor $i$ between $r_n$ and $t_n$ develops as a consequence of the restored inversion symmetry, as discussed above \eqnref{sym:inversion}.  In fact, we will later find that the phase difference between $r_{n,\hookrightarrow}$ and $r_{n,\hookleftarrow}$ is mostly immaterial (see discussion surrounding \eqnref{eqn:Z-Z_0-pathcounting}).

\section{Generalization and Symmetries\label{appendix:symmetry}}
The exact form of the $r$, $t$ and $p$ parameters in \eqnref{eqn:full-rtp} depend critically on the exact form of the interaction profile.  However, the symmetries of the system impose strong constraints on these parameters for arbitrary interaction profiles.  Again, assume no interaction for $x\leq x_n$ and $x\geq x_{n+1}$, but otherwise leave the interaction unrestricted (here the solutions will be $\propto e^{\pm ikx}$).  Then there will again be solutions.  Generically, they will satisfy conditions at $x_n$ and $x_{n+1}$.  These conditions are analogous to those calculated in Appendix \ref{appendix:barrier}.  For $x\leq x_n$ and for $x'\geq x_{n+1}$,
\begin{equation}
\braket{x}{\phi_{\uparrow\uparrow}} = \begin{bmatrix} e^{ik(x-x_{n})}\\ 0\\0\\0 \end{bmatrix},\quad\text{and}\quad
\braket{x'}{\phi_{\uparrow\uparrow}} = \begin{bmatrix} p_{n,\rightarrow} e^{ik(x'-x_{n+1})}\\ 0\\0\\0 \end{bmatrix},
\end{equation}

\begin{equation}
\braket{x}{\phi_{\downarrow\downarrow}} = \begin{bmatrix} 0\\ 0\\0\\p_{n,\leftarrow} e^{ik(x'-x_{n})} \end{bmatrix},
\quad\text{and}\quad
\braket{x'}{\phi_{\downarrow\downarrow}} = \begin{bmatrix} 0\\ 0\\0\\e^{-ik(x'-x_{n+1})} \end{bmatrix},
\end{equation}
\begin{equation}
\braket{x}{\phi_{\rightarrow}} = \begin{bmatrix} 0\\ e^{ik(x-x_{n})}\\r_{n,\hookleftarrow} e^{-ik(x-x_{n})}\\0 \end{bmatrix},
\quad\text{and}\quad
\braket{x'}{\phi_{\rightarrow}} = \begin{bmatrix} 0\\ t_{n,\rightarrow} e^{ik(x-x_{n+1})}\\0\\0 \end{bmatrix},
\end{equation}
\begin{equation}
\braket{x}{\phi_{\leftarrow}} = \begin{bmatrix} 0\\ 0\\t_{n,\leftarrow} e^{-ik(x-x_{n})}\\0 \end{bmatrix},
\quad\text{and}\quad
\braket{x'}{\phi_{\leftarrow}} = \begin{bmatrix} 0\\ r_{n,\hookrightarrow} e^{ik(x-x_{n+1})}\\e^{-ik(x-x_{n+1})}\\0 \end{bmatrix}.
\end{equation}

Constraints can be placed on $r$, $t$, and $p$ due to the symmetries of the system.  For instance, probability current conservation gives us that \begin{equation}
1=|p_{n,\leftarrow}|^2 = |r_{n,\hookrightarrow}|^2+|t_{n,\leftarrow}|^2\quad\text{and}\quad 1=|p_{n,\rightarrow}|^2 = |r_{n,\hookleftarrow}|^2+|t_{n,\rightarrow}|^2.
\end{equation}

Acting the time reversal operator $\Theta=\sigma_x^{(e)} \sigma_x^{(I)} \mathcal{K}$ (where $e$ and $I$ refer to the electronic and nuclear spins, respectively, and $\mathcal{K}$ is the complex conjugation operation) on the ``passing'' states gives
\begin{equation}
\bra{x}\Theta\ket{\phi_{\uparrow\uparrow}} = \begin{bmatrix} 0\\ 0\\0\\e^{-ik(x-x_{n})} \end{bmatrix}\quad\text{and}\quad
\bra{x'}\Theta\ket{\phi_{\uparrow\uparrow}} = \begin{bmatrix} 0\\ 0\\0\\\bar p_{n,\rightarrow} e^{-ik(x'-x_{n+1})} \end{bmatrix}.
\end{equation}
Recognizing that $\Theta \ket{\phi_{\uparrow\uparrow}}$ must be proportional to $\ket{\phi_{\downarrow\downarrow}}$, we conclude that
\begin{equation}
p_{n,\leftarrow} = (\bar p_{n,\rightarrow})^{-1}\quad\text{or}\quad p_{n,\leftarrow} = p_{n,\rightarrow}=p_n.
\end{equation}

Similarly,
\begin{equation}
\bra{x}\Theta\ket{\phi_{\rightarrow}} = \begin{bmatrix} 0\\ -\bar r_{n,\hookleftarrow} e^{ik(x-x_{n})}\\-e^{-ik(x-x_{n})}\\0 \end{bmatrix}
\quad\text{and}\quad
\bra{x'}\Theta\ket{\phi_{\rightarrow}} = \begin{bmatrix} 0\\ 0\\-\bar t_{n,\rightarrow} e^{-ik(x-x_{n+1})}\\0 \end{bmatrix}.
\end{equation}
Here, however, $\Theta\ket{\phi_{\rightarrow}}$ must be some linear combination,
\begin{equation}
\Theta\ket{\phi_{\rightarrow}}=A \ket{\phi_{\rightarrow}} +B\ket{\phi_{\leftarrow}}.
\end{equation}
The second component on the left hand side requires $A=-\bar r_{n,\hookleftarrow}$, and the third component on the right requires $B=-\bar t_{n,\rightarrow}$.  The remaining two nontrivial components give equations
\begin{equation}
-\bar r_{n,\hookleftarrow} t_{n,\rightarrow} -r_{n,\hookrightarrow} \bar t_{n,\rightarrow}= 0,\quad\text{and}\quad -|r_{n,\hookleftarrow}|^2-t_{n,\leftarrow}\bar t_{n,\rightarrow} = -1.
\end{equation}
The second equation implies
\begin{align}
1&= 1-t_{n,\rightarrow}\bar t_{n,\rightarrow} +t_{n,\leftarrow} \bar t_{n,\rightarrow} = 1-\bar t_{n,\rightarrow} \lB t_{n,\rightarrow} -t_{n,\leftarrow} \rB,
\end{align}
so that
\begin{equation}
t_{n,\rightarrow} =t_{n,\leftarrow}=t_n=|t_n|e^{i\phi_{t_n}},
\end{equation}
where the phase of $t$ has been separated out.  Separating out the phase for the reflection coefficient
\begin{equation}
r_{n,\hookleftarrow} = |r_n|e^{i\phi_{r_n}},
\end{equation}
the first equation takes the form $\bar r_{n,\hookleftarrow} t =-r_{n,\hookrightarrow} \bar t$, and therefore
\begin{equation}
r_{n,\hookrightarrow}=-e^{2i\phi_{t_n}}\bar r_{n,\hookleftarrow} =-|r_n|e^{i\lB 2\phi_{t_n}-\phi_{r_n}\rB}.
\end{equation}
Therefore, there are four independent real numbers that fully determine all the scattering amplitudes for an arbitrary potential:
\begin{itemize}
 \item the magnitude of $|r_{n,\hookleftarrow}|^2=|r_{n,\hookrightarrow}|^2=|r_n|^2$,
 \item the phase of $p_n=p_{n,\rightarrow}=p_{n,\leftarrow}$,
 \item the phase of $t_n=|t_n|e^{i\phi_{n,t}}=\sqrt{1-|r_n|^2}e^{i\phi_{t_n}}$,
 \item the phase of $r_{n,\hookleftarrow}=\sqrt{1-|t_n|^2}e^{i\phi_{r_n}}$.
\end{itemize}
The following relationship proves useful:
\begin{equation}
r_{n,\hookrightarrow} r_{n,\hookleftarrow}  = -|r_n|e^{i[2\phi_{t_n}-\phi_{r_n}]}|r_n|e^{i\phi_{r_n}}=-|r_n|^2e^{2i\phi_{t_n}}=-t_n^2\frac{|r_n|^2}{|t_n|^2}\quad\text{or}\quad \frac{r_{n,\hookrightarrow} r_{n,\hookleftarrow}}{t_n^2}= -\frac{|r_n|^2}{|t_n|^2}=\frac{|r_n|^2}{|r_n|^2-1}.\label{eqn:rt-relation}
\end{equation}

Note that if we also assume inversion symmetry in the above, $r_{n,\hookleftarrow}=r_{n,\hookrightarrow}=r$, so that $\bar r_nt_n=-r_n\bar t_n$, or $r_n\bar t_n$ is imaginary.  Or,
\begin{equation}
\phi_{t_n}=\phi_{r_n}+\pi/2.\label{sym:inversion}
\end{equation}

\section{Naive Direct Approach\label{appendix:naive-direct}}
An arbitrary state of the Hilbert space can be written as
\begin{equation}
\ket{\Psi} = \sum_{s} \int \lB \psi_{\rightarrow,s}(x)\ket{x,\uparrow}+\psi_{\leftarrow,s}(x)\ket{x,\downarrow} \rB\,dx\otimes \ket{s},
\end{equation}
where $s$ ranges over all nuclear spin configurations, and $\ket{x,\uparrow}$ and $\ket{x,\downarrow}$ are basis states on the Dirac Hilbert space for, respectively, spin-up and spin-down states.  Because of spin-momentum locking, the propagation direction of the electron automatically determines its spin; it proves more convenient to denote the electron basis states in terms of the propagation direction (using left/right arrows), so this is the notation we will use henceforth in this appendix.  For $0 < n \leq N$, define the projector
\begin{equation}
P_n = \int_{x_n}^{x_{n+1}} \lB \ket{x,\rightarrow}\bra{x,\rightarrow}+\ket{x,\leftarrow}\bra{x,\leftarrow}\rB\,dx \otimes \sum_s \ket{s}\bra{s},
\end{equation}
which projects the wavefunction to zero away from the $n$th spin.  To keep things uniform, define
\begin{equation}
P_{0} = \int_{-\infty}^{0} \lB \ket{x,\rightarrow}\bra{x,\rightarrow}+\ket{x,\leftarrow}\bra{x,\leftarrow}\rB\,dx \otimes \sum_s \ket{s}\bra{s},
\end{equation}
and
\begin{equation}
P_{N+1} = \int_{x_{N+1}}^{\infty} \lB \ket{x,\rightarrow}\bra{x,\rightarrow}+\ket{x,\leftarrow}\bra{x,\leftarrow}\rB\,dx \otimes \sum_s \ket{s}\bra{s},
\end{equation}
which, respectively, restrict wavefunctions to the left and right of the nuclear spin lattice.  Next, let $\ket{\phi^{n,(E),\text{full}}_{j}}$ be the solutions of the two-body problem:
\begin{equation}
H_0^{(n)} = -i\hbar \partial_x v \sigma_z^{(e)} + V^{(e-I_n)}(x).\label{naive-twobody}
\end{equation}
As in Appendix \ref{appendix:symmetry}, choose these eigenstates to have only rightward or leftward incoming amplitude, and incoming nuclear spin only up or down.\footnote{We are assuming there are no bound states.}  Furthermore, project these wavefunctions with $P_n$.  Using the notation from Appendix \ref{appendix:symmetry},
\begin{align}
\ket{\phi^{n,(E)}_{\leftarrow,\uparrow}}=P_n\ket{\phi^{n,(E)}_{\leftarrow}},\\
\ket{\phi^{n,(E)}_{\rightarrow,\uparrow}}=P_n\ket{\phi^{n,(E)}_{\uparrow\uparrow}},\\
\ket{\phi^{n,(E)}_{\leftarrow,\downarrow}}=P_n\ket{\phi^{n,(E)}_{\downarrow\downarrow}},\\
\ket{\phi^{n,(E)}_{\rightarrow,\downarrow}}=P_n\ket{\phi^{n,(E)}_{\rightarrow}}.
\end{align}
Similarly, let the $n=0$ and $n=N+1$ states \begin{equation}
\ket{\phi^{n,(E)}_{\leftarrow}},\quad\text{and} \quad \ket{\phi^{n,(E)}_{\rightarrow}},
\end{equation}
refer to the plane wave solutions (i.e., for $H=\hbar v\sigma_x^{(e)}$) restricted, respectively, to the left and right of the nuclear spin lattice (i.e., acted on by $P_0$ and $P_{N+1}$, respectively).  Notice that these states together span the full Hilbert space---but are not a basis; they are overcomplete.  Indeed, every energy eigenstate (of energy $E$) can be written as
\def\nslash{{\slash\hspace{-0.42em}n}}
\begin{multline}
\ket{\Psi} = \sum_{n=1}^{N} \sum_{s_\nslash}\lB \alpha_{n,s_\nslash} \ket{\phi^{n,(E)}_{\rightarrow,\uparrow}}\ket{s_\nslash} + \beta_{n,s_\nslash} \ket{\phi^{n,(E)}_{\leftarrow,\uparrow}}\ket{s_\nslash} + \alpha'_{n,s_\nslash} \ket{\phi^{n,(E)}_{\rightarrow,\downarrow}}\ket{s_\nslash} +\beta'_{n,s_\nslash} \ket{\phi^{n,(E)}_{\leftarrow,\downarrow}}\ket{s_\nslash} \rB\\
 + \sum_{s}\lB \zeta_{-1,s}\ket{\phi^{-1,(E)}_{\rightarrow}}\ket{s} + \zeta_{-1,s}' \ket{\phi^{-1,(E)}_{\leftarrow}}\ket{s} + \zeta_{N+1,s}\ket{\phi^{N+1,(E)}_{\rightarrow}}\ket{s} + \zeta_{N+1,s}' \ket{\phi^{N+1,(E)}_{\leftarrow}}\ket{s}\rB,
\end{multline}
where $s_\nslash$ is the spin configuration for all nuclear spins other than $n$, and $\ket{s_\nslash}'$ the corresponding vector in $\mathcal{H}_I^{\otimes N-1}$.  Notice that not every combination is a energy eigenstate: if appropriate boundary conditions are not met, jump discontinuities exist, and the derivative in \eqnref{naive-twobody} then leads to Dirac delta functions.

The approach for the single nucleus case followed the basic structure:
\begin{enumerate}
 \item Find solutions in particular regions
 \item Match boundary conditions
\end{enumerate}

We have already described the many-body system in terms of solutions in ``particular regions,'' though now we have $2^N$ ``spin regions'' for each $N+2$ spatial regions---for a total size of $(N+2)2^N$ regions.  The size of this parameter space is what makes this problem difficult.

\subsection{Boundary Matching}
Consider the case where an electron is injected on the left, moving right, into an initial nuclear spin configuration $s$.  In this case, the boundary conditions fully specify the scattering eigenstate:
\begin{multline}
\ket{\Psi_{\rightarrow e,s}^{(E)}} = \overbrace{\sum_{n=1}^{N} \sum_{s_\nslash}\lB \alpha_{n,s_\nslash} \ket{\phi^{n,(E)}_{\rightarrow,\uparrow}}\ket{s_\nslash}' + \beta_{n,s_\nslash} \ket{\phi^{n,(E)}_{\leftarrow,\uparrow}}\ket{s_\nslash}' + \alpha'_{n,s_\nslash} \ket{\phi^{n,(E)}_{\rightarrow,\downarrow}}\ket{s_\nslash}' +\beta'_{n,s_\nslash} \ket{\phi^{n,(E)}_{\leftarrow,\downarrow}}\ket{s_\nslash}' \rB}^\text{internal degrees of freedom}\\
 + \overbrace{\ket{\phi^{0,(E)}_{\rightarrow}}\ket{s}}^\text{incoming wave} + \overbrace{\sum_{s'} \lB \zeta_{0,s'}' \ket{\phi^{0,(E)}_{\leftarrow}}\ket{s'} + \zeta_{N+1,s'}\ket{\phi^{N+1,(E)}_{\rightarrow}}\ket{s'}\rB}^\text{outgoing wave}.
\end{multline}
Moreover, notice that each boundary matching conserves total spin.  This allows for some simplifications:
\begin{equation}
\ket{\Psi_{\rightarrow e,s}^{(E)}} = \ket{\phi^{-1,(E)}_{\rightarrow}}\ket{s} + \sum_{n=0}^{N-1} \sum_{s'} A_{n,s'} \ket{\phi^{n,(E)}_{\gamma_{s,s'},s_n}}\ket{s'}' + \sum_{s'} \mathcal{Z}_{s,s'} \ket{\phi^{\gamma_{s,s'}',(E)}_{\gamma_{s,s'}}}\ket{s'},
\end{equation}
where $\gamma_{s,s'}$ is $\rightarrow$ if $J_z[s']=J_z[s]$, and $\leftarrow$ if $J_z[s']=J_z[s]+\hbar$ (and does not matter for other values).  Similarly, $\gamma'$ is, respectively $N+1$ or $0$.  Note also that, by conservation of probability current, the total incoming amplitude equals the sum of the total outgoing amplitudes:
\begin{equation}
1 = \sum_{s'} \lA \mathcal{Z}_{s,s'}\rA^2.
\end{equation}

The direct approach to satisfying all the boundary conditions is to consider all possible ``paths'' of the electron through the nuclear spin lattice.  The word is placed in quotes because here we are using it to refer not to a physical path---quantum or classical---but rather to the \emph{procedure} one follows to satisfy all boundary conditions by choosing appropriate wavefunction amplitudes.  It is important to stress that, while these paths can actually be related to some interpretation of the electron's actual path, what is being referred to here is simply the mathematical procedure to satisfy all boundary conditions.  Each ``path'' $\pi$ consists of a sequence of adjacent sites $1\leq \pi_i\leq N$ (with $\pi_0=0$, and $\pi_\text{final}$ either $0$ or $N+1$), and with certain constraints placed on the wavefunction at each site.  For example, $\pi=[010]$ corresponds to an electron coming from the left, reflecting off of the first site and exiting to the left.

To help understand this process, consider an electron incoming from the left and impinging on the nuclear spin configuration $\uparrow\uparrow\downarrow$ as an example.  Here, we assume that all the single-nucleus scattering amplitudes are independent of the site index $n$ and that each interaction profile is inversion symmetric for simplicity.  We will match boundary conditions.  By definition, the far left, right-moving segment must have coefficient $1$.
\begin{center}
\includegraphics{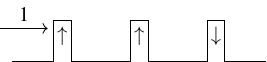}
\end{center}
Because each boundary matching step does not change the total angular momentum, all left moving electronic wavefunctions vanish.  Moreover, all right moving wavefunctions are easily seen to collect either a factor of $p$ (from both of the first two spins), or $t$ (from the last spin).
\begin{center}
\includegraphics{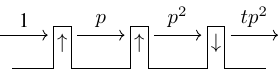}
\end{center}
This process corresponds to the path $\pi=[01234]$, or
\begin{equation}
\pi(0)=0;\quad \pi(1)=1;\quad \pi(2)=2;\quad \pi(3)=3;\quad \pi(4)=4.
\end{equation}
Next, lets find the wavefunction in the spin sector $\uparrow\uparrow\uparrow$.  The only important boundary condition for this sector is the reflection off site $3$.  We already calculated that in the above, so it is easy to find the wave function,
\begin{center}
\includegraphics{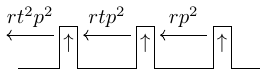}
\end{center}
and, as above, all right-moving components vanish.  This process used information about the wavefunction calculated up to the third step of $\pi$, and then worked back to the far left hand side: $\pi'=[0123210]$.  Repeat the process for $\uparrow\downarrow\uparrow$ and $\downarrow\uparrow\uparrow$.
\begin{center}
\includegraphics{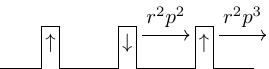}
\end{center}
\begin{center}
\includegraphics{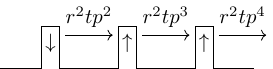}
\end{center}
These two processes corresponded to paths $[0123234]$ and $[012321234]$, respectively.  Notice that any other path will provide no new information about the wavefunction: every other path would involve another reflection, but that would have amplitude zero.  Moreover, notice that at any step of the path, the state of the spin sector (i.e., the spin configuration that is matched) had some value, $s_{\pi; i}(n)$.  We will often use heuristic language, referring to the electron as moving from site $\pi(i-1)$ to $\pi(i)$, and the spin configuration is in the state $s_{\pi;i}$---even though it is not strictly correct.  In this same sense, the spin must move right as you follow a path (taking into account the electrons spin).  It follows that there are finitely many paths to consider to match all boundary conditions.

Moreover, there may be multiple paths that impose requirements on the same part of the wavefunction.  Because the boundary conditions impose linear constraints, the requirements for different paths are simply summed.  The final outgoing amplitude for the nuclear spin lattice to start in configuration $\mathbf{m}$ and end in configuration $\mathbf{m}'$ can be expressed in terms of paths:
\begin{equation}
\mathcal{Z}_{\mathbf{m},\mathbf{m}'} = \sum_{\text{paths $\pi$ producing $\mathbf{m}'$}} \prod_n r_{n,\hookleftarrow}^{N_{r_{\hookleftarrow}}(\pi;n)}r_{n,\hookrightarrow}^{N_{r_{\hookrightarrow}}(\pi;n)}t_n^{N_t(\pi;n)}p_n^{N_p(\pi;n)}\label{eqn:ZZ-def}
\end{equation}
where $N_{r_{\hookrightarrow}}(\pi;n)$ is the number of reflections (outgoing to the right) in path $\pi$ at site $n$, and so on.
\subsubsection*{Example Calculation}
A more complicated example is included here, fully completed.  Once understood, the process is purely mechanical.
\begin{center}
\includegraphics{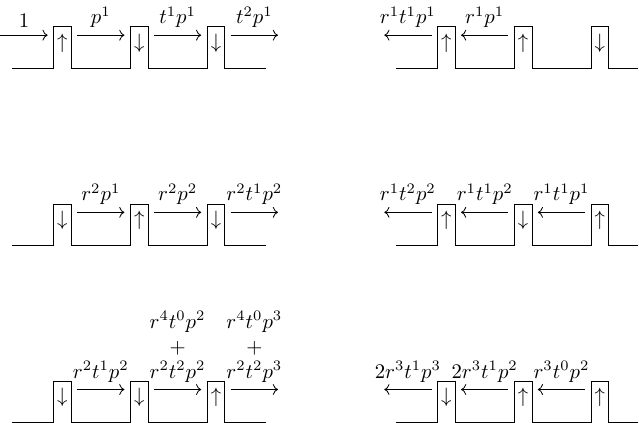}
\end{center}

\subsection{Naive Algorithm}
As seen from the example, an algorithm can be followed.
\begin{enumerate}
 \item Fix an initial spin configuration.  Put the electron on the far left, moving right (up-spin).  Add this initial configuration to a list of configurations to consider.  \item While there are still configurations in this list, process each configuration as follows:
\begin{enumerate}
 \item If the electron is moving right and at the far right, or moving left and at the far left, this is a terminal configuration.  Record it.
 \item If the electron is moving right (left) and the right (left) spin is up (down), the electron simply passes through the barrier.  Add this same configuration, but with one additional $p_n$ multiplying the amplitude.
 \item If the electron is moving right (left) and the right (left) spin is down (up): the electron can either transmit through the barrier (with an amplitude multiplied by $t_n$) add this to the list of configurations to consider.  Or, the aforementioned spin can flip, the electron can flip direction, and the amplitude can be multiplied by $r_{n,\hookleftarrow}$ ($r_{n,\hookrightarrow}$).  Add this to the list of configurations to consider.
\end{enumerate}
\item At the end, each terminal state may be recorded multiple times---representing different paths through the lattice.  Add all the possible amplitudes---this is the final amplitude for that configuration.
\end{enumerate}

This algorithm must take time proportional to the number of paths---which grows very quickly with system size.  If the individual paths are not needed (i.e., if the only the outgoing amplitudes are desired), there are simplifications.  Appendix \ref{appendix:spin-migration} addresses this.

\section{Combinatorial Reduction\label{appendix:spin-migration}}
The high symmetry of the situation permits a dramatic reduction in the complexity of the calculation.  In particular, there are several nontrivial restrictions on the number of reflections, passes, and transmissions of any permitted ``path.''  Here, these restrictions are identified, and then used to reduce the problem to a much simpler combinatorial problem.

\subsection{Spin Migration: relationships between $N_r$, $N_t$, and $N_p$}
In this section, the fact that the system conserves spin will be exploited to derive some useful relationships.  First, define $\delta J$ to be the (normalized) change in the nuclear spin on site $n$:
\begin{equation}
\delta J_{\mathbf{m},\mathbf{m}'}(n) = \frac{\mathbf{m}_i'-\mathbf{m}_i}{2} = N_{r_{\hookleftarrow}}(n)-N_{r_{\hookrightarrow}}(n),\label{eqn:site-torque}
\end{equation}
where $\mathbf{m}_n$ is $\pm1$ for spin-up and spin-down, $\mathbf{m}$ denotes the input spin configuration, and $\mathbf{m}'$ the final configuration.  We have also included an observation that each reflection to the left (right), $N_{r_{\hookleftarrow}}(n)$ ($N_{r_{\hookrightarrow}}(n)$) imparts $+2$ ($-2$)
angular momentum to the site.  Consider the (normalized) net change in spin in the first $n$ sites:
\begin{equation}
\Delta J_{\mathbf{m},\mathbf{m}'}(n) = \sum_{i\leq n}\delta J_{\mathbf{m},\mathbf{m}'}(n).
\end{equation}
Since spin must move right in this system, and exactly one spin is injected at the far left, this quantity is at most $1$.  By spin conservation, exactly $1-\Delta J_{\mathbf{m},\mathbf{m}'}(n)$ up spins must be deposited on the sites right of $n$ or exit on the right.  This means that an up spin must move
right from site $n$ to site $n+1$ (or exit right) exactly
\begin{equation}
N_{p_\rightarrow}(n)+N_{t_\rightarrow}(n)+N_{r_{\hookrightarrow}}(n)
= N_{p_\rightarrow}(n+1)+N_{t_\rightarrow}(n+1)+N_{r_{\hookleftarrow}}(n+1)
= \lambda_{\mathbf{m},\mathbf{m}'}(n) = 1-\Delta J_{\mathbf{m},\mathbf{m}'}(n)
\end{equation}
times.  $N_{p_\rightarrow}(n')$ is the number of passes (of an electron moving rightward) at site $n'$, and likewise for the other symbols; the expression should be ignored for $n'=0,N+1$.  Similarly, since the electron is neither created nor destroyed, it must move left between these same sites
\begin{equation}
N_{p_\leftarrow}(n)+N_{t_\leftarrow}(n)+N_{r_{\hookrightarrow}}(n)
= N_{p_\leftarrow}(n+1)+N_{t_\leftarrow}(n+1)+N_{r_{\hookleftarrow}}(n+1)
= \rho_{\mathbf{m},\mathbf{m}'}(n)
= \fun 1-\Delta J_{\mathbf{m},\mathbf{m}'}(n),& \text{left exiting}\\ \hphantom{1}-\Delta J_{\mathbf{m},\mathbf{m}'}(n),& \text{right exiting}\fund .
\end{equation}
It follows then, that the total number of interactions at a site is
\begin{equation}
N_t(n)+N_p(n)+N_r(n)=\lambda_{\mathbf{m},\mathbf{m}'}(n-1)+\rho_{\mathbf{m},\mathbf{m}'}(n)
= \fun 2-2\Delta J_{\mathbf{m},\mathbf{m}'}(n-1)-\delta J_{\mathbf{m},\mathbf{m}'}(n),& \text{left exiting}\\
1-2\Delta J_{\mathbf{m},\mathbf{m}'}(n-1)-\delta J_{\mathbf{m},\mathbf{m}'}(n),& \text{right exiting}\fund.\label{eqn:total-interaction-count}
\end{equation}

To visualize the situation, consider a site $n$ and initial and final configuration $\mathbf{m}$ and $\mathbf{m}'$, respectively.  By the discussion above, we know how many approaches and departures are made from the right and left to this site---numbers independent of the path.  Label each approach and each departure on each side---i.e., left approach \#1, left departure \#1, etc (call these sequence numbers).  Every path is uniquely identified by specifying where each arrival must go (i.e., reflecting or not).  Illustrations are provided below, for left-departing electrons (for $\mathbf{m}_n=-1,-1,1,1$, and $\mathbf{m}'_n=-1,1,1,-1$),
\begin{center}
\includegraphics{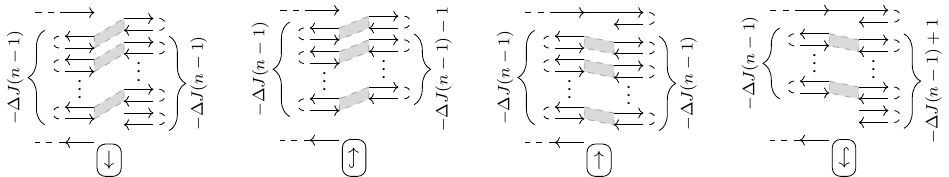}
\end{center}
and right-departing electrons.
\begin{center}
\includegraphics{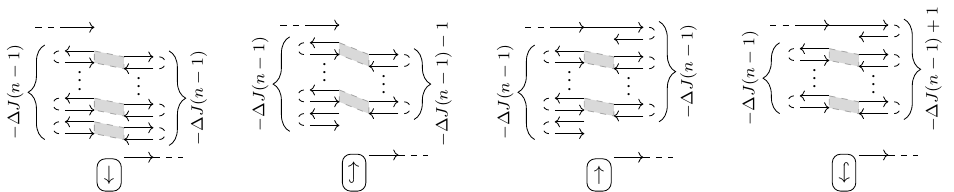}
\end{center}

The above illustrates the situation for all 4 combinations of initial and final state, for both left and right exiting cases (8 total), and constitutes a complete case-breakdown.  The right and left facing solid arrows represent the right and left approaches to the nuclear spin (visualized as the circled arrow, which graphically indicates its initial and final state).  The ordering in the vertical direction indicates the sequence number (downward is increasing), and is enforced by requiring that connections do not skip over yet-unused sequence numbers.  The dashed semicircles to the far left and right indicate some combination of reflections occurring elsewhere in the lattice.  Some arrows are shown already connected---indicating that all valid paths have this connection.  The remaining unconnected arrow heads and tails must be connected, such that the ordering of the arrows in the path is maintained (and that a reflection is allowed by the spin state at the site).  The grey boxes connecting quadruples of arrows can be used here to guide the eye, visually grouping the terms (but will be used later).

For example, for the left exiting, $\mathbf{m}_n=\mathbf{m}'_n=-1$ case, the initial arrow has two choices: right exiting \#1 (transmission), or left exiting \#2 (reflection).  Any other connection violates the assertion that the path segments are already sequenced.  Compare this to the illustration for $\mathbf{m}_n=\mathbf{m}'_n=1$: here left entering \#1 has only one choice to connect: right exiting \#1 (a pass), because a reflection would not be possible.  That connection is already made in the illustration.

After one is convinced of the validity of this case breakdown, it is clear that each transmission or reflection must be followed by a pass: if another approach occurs, the spin cannot have changed.  In fact, by considering each of the cases above, we conclude that
\begin{equation}
N_p(n) = \fun N_t(n) + N_r(n)-\delta J_{\mathbf{m},\mathbf{m}'}(n), & \text{left exiting} \\
N_t(n) + N_r(n)+\delta J_{\mathbf{m},\mathbf{m}'}(n) + \mathbf{m}_n, & \text{right exiting}\fund.\label{eqn:delta_0-identity}
\end{equation}
Combining this with \eqnref{eqn:total-interaction-count}, we obtain
\begin{equation}
N_p(n) = \fun 1-\Delta J_{\mathbf{m},\mathbf{m}'}(n), & \text{left exiting} \\
1-\Delta J_{\mathbf{m},\mathbf{m}'}(n-1)+\frac{\mathbf{m}_n-1}{2}, & \text{right exiting}\fund,\label{eqn:n_p}
\end{equation}
and a quantity we call the ``surplus,'' that will prove critically important:
\begin{equation}
 N^0_{\mathbf{m},\mathbf{m}'}(n)=N_t(n)+N_r(n)=\fun 1-\Delta J_{\mathbf{m},\mathbf{m}'}(n-1),& \text{left exiting}\\ 1-\Delta J_{\mathbf{m},\mathbf{m}'}(n-1)-(\mathbf{m}'(n)+1)/2,&\text{right exiting}\fund.\label{def:surplus}
\end{equation}

Summarizing the results, the number of passes $N_p(n)$ at each site is determined by the initial and final nuclear spin configurations; if the number of reflections at each site $N_r(n)$ is also specified, so is the number of transmissions (as well as the number of rightward and leftward reflections).    \eqnref{eqn:ZZ-def} can be rewritten by summing over the possible numbers of reflections at each site $N_r(n)$, and introducing the path-counting function $\mathcal P_{\mathbf{m},\mathbf{m}'}(N_r)$, which we will explore momentarily:
\begin{align}
\mathcal{Z}_{\mathbf{m},\mathbf{m}'}
&= \sum_{\lC N_r(n)\rC} \mathcal P_{\mathbf{m},\mathbf{m}'}(N_r) \prod_n r_{n,\hookleftarrow}^{N_{r_{\hookleftarrow}}(n)}r_{n,\hookrightarrow}^{N_{r_{\hookrightarrow}}(n)}t_n^{N_t(n)}p_n^{N_p(n)} \nonumber\\
&= \lB \prod_n p_n^{N_p(n)} t_n^{ N^0(n)} \rB
\sum_{\lC N_r(n)\rC} \mathcal P_{\mathbf{m},\mathbf{m}'}(N_r) \prod_n(r_{n,\hookleftarrow}/t_n)^{N_{r_{\hookleftarrow}}(n)}(r_{n,\hookrightarrow}/t_n)^{N_{r_{\hookrightarrow}}(n)} \nonumber\\
&= \lB \prod_n p_n^{N_p(n)} t_n^{ N^0(n)} \rB \sum_{\lC N_r(n)\rC} \mathcal P_{\mathbf{m},\mathbf{m}'}(N_r) \prod_n e^{i\lB  N_{n,r_{\hookleftarrow}}(\phi_{r_n}-\phi_{t_n}) + N_{n,r_{\hookrightarrow}}(\phi_{t_n} - \phi_{r_n}+\pi) \rB} |r_n/t_n|^{N_{r}(n)}\nonumber\\
&= \overbrace{\lB \prod_n p_n^{N_p(n)} t_n^{ N^0(n)} e^{i(\phi_{r_n}-\phi_{t_n}-\pi/2)\delta J(n)}  \rB}^{\mathcal{Z}_{0;\mathbf{m},\mathbf{m}'}} \sum_{\lC N_r(n)\rC} \mathcal P_{\mathbf{m},\mathbf{m}'}(N_r) \prod_n  (\overbrace{i|r_n/t_n|}^{z_n}){}^{N_{r}(n)}
,\label{eqn:Z-Z_0-pathcounting}
\end{align}
where we have used the fact that
$N_{r_{\hookleftarrow}}(n)-N_{r_{\hookrightarrow}}(n)=\delta J(n)$.  When $\delta J(n)=0$, \eqnref{eqn:rt-relation} cancels phases associated with reflection.  The residual phase when $\delta J(n)=-1$ combines with the factor of $i$ (because $N_r$ must be odd) to give the extra minus $\pi$ phase from $r_{n\hookrightarrow}$.

\subsection{The ``Surplus'' $ N^0_{s,s'}(n)$}
Finally, a combinatorial argument is used to count all paths that (a) produce a particular final spin configuration $\mathbf{m}'$, and (b) has a specified number of reflections at each site, $N'_r(n)$.  The formula is surprisingly compact:
\begin{equation}
\mathcal P_{\mathbf{m},\mathbf{m}';\lC N'_r(n)\rC } = \prod_n { N^0_{\mathbf{m},\mathbf{m}'}(n) \choose N'_r(n)},\label{paths-byreflection}
\end{equation}
To break down and prove this relationship, first we show that the number of paths can be expressed as a product over data about each site.  To see this,
notice that, in the above case breakdown, if two paths agree on all of these connections, on all sites, the paths are the same.  Similarly, if the paths differ for any choice of connection at any site, they are different.  It follows that the number of paths is therefore a product of the number of ways of connecting these arrows at each site.

It remains to be proven that the number of ways to connect these arrows is precisely ${ N^0_{\mathbf{m},\mathbf{m}'}(n) \choose N'_r(n)}$.  This follows from a simple, albeit tedious, case analysis.  The above visualization makes this verification relatively easy: observe that each shaded block connects 4 segments.  The interaction is simple: a choice of upper, outgoing arrow (either right or left) is made; once done, the electron must come back on the respective incoming segment; the spin at this point only allows the electron to then pass through to the unchosen outgoing segment, and then brought back in along the final, incoming segment.  A single choice must be made (to either reflect or not) in each box.  Carefully tracking through all the blocks shows there is a single, additional possible location for reflection at the very end.  Counting these up, we find that the surplus, as defined, properly captures the counting factor.

\subsection{Polynomial Form}
Working with these products is vastly simpler than the direct counting described in Appendix \ref{appendix:naive-direct}, but there are still some meaningful simplifications.  The total number of paths with $N_r$ reflections is
\begin{equation}
\mathcal P_{\mathbf{m},\mathbf{m}'}(N_r)=\sum_{\lC N'_r(n)\middle| \sum_n N'_r(n)=
N_r\rC}  \prod_n {  N^0_{\mathbf{m},\mathbf{m}'}(n) \choose N_r'(n)}.
\end{equation}
The $N_r(n)'$ are restricted here to be valid for the $\mathbf{m}$ to $\mathbf{m}'$ transformation.  Using \eqnref{eqn:Z-Z_0-pathcounting}, we have
\begin{equation}
\mathcal{Z}_{\mathbf{m},\mathbf{m}'} = \mathcal{Z}_{0;\mathbf{m},\mathbf{m}'} \sum_{\lC N'_r(n)\rC} \prod_n {  N^0_{\mathbf{m},\mathbf{m}'}(n) \choose N_r'(n) } z_n^{N'_r(n)}.
\end{equation}
Furthermore, the restriction on $N'_r(n)$ is still implied, but can be expressed simply in terms of the absolute value of $\delta J_{\mathbf{m},\mathbf{m}'}(n)$:
\begin{equation}
\mathcal{Z}_{\mathbf{m},\mathbf{m}'} = \mathcal{Z}_{0;\mathbf{m},\mathbf{m}'} \sum_{\lC k_n\rC} \prod_n {  N^0_{\mathbf{m},\mathbf{m}'}(n) \choose |\delta J_{\mathbf{m},\mathbf{m}'}(n)| +2k_n} z_n^{|\delta J_{\mathbf{m},\mathbf{m}'}(n)| +2k_n}.
\end{equation}
This allows an interchange of sum and product, reducing the calculation to a polynomial multiplication problem:
\begin{equation}
\mathcal{Z}_{\mathbf{m},\mathbf{m}'} = \mathcal{Z}_{0;\mathbf{m},\mathbf{m}'} \prod_n \sum_{k}  {  N^0_{\mathbf{m},\mathbf{m}'}(n) \choose |\delta J_{\mathbf{m},\mathbf{m}'}(n)| + 2k} z_n^{|\delta J_{\mathbf{m},\mathbf{m}'}(n)| +2k}. \label{paths-Z}
\end{equation}
Alternatively, the sum over $k$ can be expressed in terms of binomial factors:
\begin{align}
\mathcal{Z}_{\mathbf{m},\mathbf{m}'} &= \mathcal{Z}_{0;\mathbf{m},\mathbf{m}'}\prod_n \frac{1}{2}\lP (1+z_n)^{ N^0_{\mathbf{m},\mathbf{m}'}(n)} + (-1)^{|\delta J_{\mathbf{m},\mathbf{m}'}(n)|} (1-z_n)^{ N^0_{\mathbf{m},\mathbf{m}'}(n)} \rP \nonumber\\
&= \frac{\mathcal{Z}_{0;\mathbf{m},\mathbf{m}'}}{2^{N}} \prod_n \lP (1+z_n)^{ N^0_{\mathbf{m},\mathbf{m}'}(n)} + (-1)^{|\delta J_{\mathbf{m},\mathbf{m}'}(n)|} (1-z_n)^{ N^0_{\mathbf{m},\mathbf{m}'}(n)} \rP.
\end{align}

\section{Total Spin Movement\label{appendix:misc}}
Define the change in ``center of angular momentum'' (measured from the far right),
\begin{equation}
d_S(\mathbf{m},\mathbf{m}')
=\sum_n (N+1-n)\mathbf{m}_n'-\sum_n (N+1-n)\mathbf{m}_n
= \sum_n (N+1-n)\delta J_{\mathbf{m},\mathbf{m'}}(n)
= \sum_n \Delta J_{\mathbf{m},\mathbf{m'}}(n)
. \label{def:d_S}
\end{equation}
Summing \eqnref{eqn:n_p} over all sites,
\begin{equation}
N_p = \fun N-d_S(\mathbf{m},\mathbf{m}'), & \text{left exiting} \\
N-d_S(\mathbf{m},\mathbf{m}')+\frac{N_\uparrow-N}{2}, & \text{right exiting}\fund.
\end{equation}

\end{document}